\documentclass[12pt]{iopart}

\newcommand{\eq}[1]{\begin{equation}#1\end{equation}}

\usepackage{graphics}
\usepackage{graphicx}
\usepackage{enumerate}

\begin{document}

\title{Entanglement evolution after connecting finite to infinite
quantum chains}
\author{V. Eisler$^1$, D. Karevski$^2$, T. Platini$^2$ and I. Peschel$^1$}
\address{$^1$ Fachbereich Physik, Freie Universit\"at Berlin, Arnimallee 14,
D-14195 Berlin, Germany\\
$^2$ Laboratoire de Physique des Mat\'eriaux, UMR CNRS 7556, Universit\'e Henri
Poincar\'e, Nancy 1, B.P. 239, F-54506 Vandoeuvre-les-Nancy Cedex, France}


\begin{abstract}
We study zero-temperature XX chains and transverse Ising chains and join an initially 
separate finite piece on one or on both sides to an infinite remainder. 
In both critical and non-critical systems we find a typical
increase of the entanglement entropy after the quench, followed by
a slow decay towards the value of the homogeneous chain. In the 
critical case, the predictions of conformal field theory are verified 
for the first phase of the evolution, while at late times a step structure
can be observed. 
\end{abstract}

\section{Introduction}

The complexity of many-body quantum states manifests itself in the connection
between different parts of a system in the wave function. It can be
measured by the entanglement entropy $S$ which has been studied extensively
over the last years and shows, for example, a logarithmic divergence
at quantum phase transitions in chains \cite{Amicoetal07}. More recently,
its temporal evolution for evolving quantum states has come into the focus
of the studies. The simplest situation is a quench, where the Hamiltonian is
changed instantaneously from $H_0$ to $H_1$ and one follows the subsequent
evolution of the initial state, usually taken to be the ground state of $H_0$.
The first such studies considered global quenches, in which a parameter is
changed everywhere in the same way \cite{CC05,DeChiara06}. It was found that
in this case the entanglement entropy becomes extensive, in contrast to
the equilibrium states where it is proportional to the interface area between
the two chosen parts of the system.

A different situation is that of a local quench, where a parameter is changed
only in one or in a few places. For example, a single nearest-neighbour bond 
could be modified. In one dimension, this can have strong effects, since by 
removing or adding the full bond, one cuts the chain or joins previously
separate pieces. In the first case, the initial entanglement is conserved,
but in the second case it shows interesting temporal behaviour. This was
investigated recently in a numerical study of free electrons hopping on a
chain with initially one defect in form of a weakened bond which was then 
removed
\cite{EislerPeschel07}. In spin language, this corresponds to an XX model.
The entanglement entropy was found to increase to a 
maximum, or plateau, followed by a slow decay to its equilibrium value.
The maximum value was proportional to $\ln L$, where $L$ is the size of
the chosen subsystem, and the whole plateau could be described by a 
logarithmic expression containing $L$ and the elapsed time $t$. In a 
subsequent publication, Calabrese and Cardy \cite{CC07} showed how to 
obtain such plateaus and the corresponding formulae from conformal
field theory. They also found that the long-time tails in $S$ are absent
in this continuum treatment and must therefore be lattice effects.

In the present study we continue these investigations by looking at chains 
where a finite segment is initially fully separated from the rest and then 
joined to it. This can be done on one or on both sides of the segment
and the rest is assumed to be infinite. All parts of the total system are
initially in their ground states. As examples we consider the XX model and
the transverse Ising (TI) model for which the entanglement can be obtained in 
a relatively simple way from the correlation functions. This is outlined in 
the following section 2. The critical case is then studied in section 3. 
As for the single defect, we find logarithmic plateaus for $S(t)$ which 
turn out to be in very good agreement with conformal predictions. For the
doubly-infinite geometry we derive the corresponding formula by extending 
the treatment in \cite{CC07}. We also illustrate how the entanglement process 
manifests itself via fronts in the eigenvectors of the reduced density matrix 
and show the evolution of the single-particle spectra underlying $S$. 
An interesting new phenomenon appears in the long-time behaviour of $S(t)$, 
where one finds a step structure in the slow decay towards equilibrium. The 
non-critical case is investigated for the TI model in Section 4. Here we find 
a plateau in $S$ as well, but with a flat shape (up to oscillations) and
a limited height depending on the transverse field. Our findings are summarized
in Section 5. Two appendices treat the relation between the entanglement entropies
of the XX and the critical TI model and the derivation of the conformal formula
for one of the geometries.

\section{Models and geometries }

In the following we study the XY quantum chain with  Hamiltonian
\eq{
 H=- \frac {1}{2} \sum_{n=1}^{N-1} \left[ \frac{1+\kappa}{2}\sigma_n^x 
\sigma_{n+1}^x+\frac{1-\kappa}{2}\sigma_n^y \sigma_{n+1}^y\right]
		- \frac {h}{2} \sum_{n=1}^{N} \sigma_{n}^z,
\label{eq:Hamiltonian}
}
where the $\sigma_n^\alpha$ are Pauli matrices, for two special cases, namely
%
\begin{itemize}

\item $\kappa=h=0$ which is the XX chain, critical and equivalent to a
   half-filled electronic hopping model

\item $\kappa=1$ which is the Ising chain in a transverse field.

\end{itemize}
%
\par
As is well known, the operator $H$ becomes a quadratic form when expressed
in terms of Fermi operators $c_n,c_n^{\dag}$ and reads after diagonalization
\eq{
H=\sum_q \omega_q\left(\eta_{q}^{\dag}\eta_{q}-1/2\right)
\label{eq:Hdiag}
}
with
\eq{
\omega_q=\sqrt{\kappa^2\sin^2q+(h+\cos q)^2}\; .
\label{eq:dispersion}
}
In general, the $\eta_{q}$ are obtained from the $c_n,c_n^{\dag}$ 
via a Bogoliubov transformation with two sets $\phi_q(n),\psi_q(n)$ of coefficients
\cite{LSM}. For the XX case this reduces to a simple linear combination of
the $c_n$ and one can also write $H$ as
\eq{
H_{XX}=- \sum_q \cos q \,\eta_{q}^{\dag}\eta_{q}
\label{eq:HamiltonianXX}
}
with $q= \pi n /(N+1), n= 1,2,...N$ for open boundary conditions.
\par
\begin{figure}[thb]
\center
\includegraphics[scale=.5]{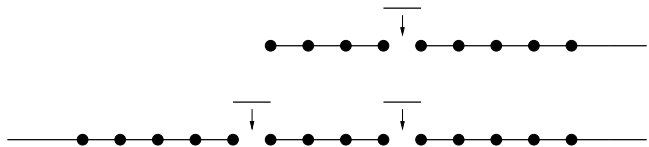}
\caption{The two geometries studied in the paper.}
\label{fig:geometry}
\end{figure}

We will investigate the two situations depicted in Fig. \ref{fig:geometry}. In the first case the 
system consists 
initially of a segment of length $L$ and a rest of length $N-L$ where $N \gg L$, both of which are
in their ground state. We will call this the semi-infinite geometry. In the second case, the 
segment is sandwiched between two such environments. This will be called the infinite geometry.
At time $t=0$, the missing bond(s) will be added and the time evolution of the state monitored.
\par
%
%
We will be interested in the entanglement of the segment with the rest of the system.
The corresponding entanglement entropy $S$ follows from the reduced density matrix $\rho$ of 
the segment which has the form \cite{Peschel03}
\eq{
\rho = \frac{1}{Z}\; e^{-\cal{H}} \; , \quad
{\cal{H}} = \sum_{k=1}^{L} \varepsilon_k(t) f_k^{\dagger} f_k 
\label{eq:rho}
}
Here $Z$ is a normalization constant ensuring $\mathrm{Tr}\,\rho = 1$ 
and the fermionic operators $f_k$ follow from the $c_n$ by an orthogonal
transformation. Then $S= -\mathrm{Tr}\,(\rho \ln \rho)$ is determined by the single-particle
eigenvalues $\varepsilon_k(t)$ according to
\eq{
S(t)=-\sum_k \zeta_k(t) \ln \zeta_k(t) - 
\sum_k (1-\zeta_k(t)) \ln (1-\zeta_k(t)),
\label{eq:entropy}
}
where $\zeta_k(t)=1/(\exp(\varepsilon_k(t))+1)$.
\par
Since the initial state is a Slater determinant, the $\varepsilon_k(t)$ resp. the  $\zeta_k(t)$
follow from the one-particle correlation functions at time $t$. These can be combined into a 
$2N \times 2N$ matrix
\begin{eqnarray} 
	\mathbf{G}=\left(\begin{array}{cc}
	\langle \Gamma_m^1 \Gamma_n^1\rangle & \langle \Gamma_m^1 \Gamma_n^2\rangle \\ 
        \langle \Gamma_m^2 \Gamma_n^1\rangle & \langle \Gamma_m^2 \Gamma_n^2\rangle	 
	\end{array}\right)
\label{eq:matrixG}
\end{eqnarray}
where $\Gamma_n^1 =(c_n^\dag+c_n)$ and $\Gamma_n^2=-i(c_n^\dag-c_n)$. Restricting this matrix
to the $L$ sites in the subsystem, its eigenvalues are $\zeta_k$ and $1-\zeta_k$. In the XX case,
it is sufficient to consider the $N \times N$ correlation matrix
\eq{
{\bf{C}}= \langle c_m^\dag c_n \rangle 
\label{eq:matrixC}
}
which has eigenvalues $\zeta_k$ when restricted to the subsystem.
\par
The time evolution of the fermionic operators with the final Hamiltonian gives the correlation
matrix at time $t$ in terms of its initial value as \cite{IgloiRieger,Karevski02}
\eq{
{\bf{G}}(t)={\bf{R}}(t) {\bf{G}}(0) {\bf{R}}^T(t)
\label{eq:evolutionG}
}
where ${\bf{R}}(t)$ is a rotation matrix with elements which are sums containing the  
$\phi_q(n),\psi_q(n)$ and exponential factors $\exp(i\omega_q t)$. In the
critical case and taking the thermodynamic limit, they can be expressed in terms
of Bessel functions. The initial correlations are particularly simple
for the XX model, where one finds for a segment with sites between $1$ and $L$ 
\eq{
C_{ij}=  \frac{1}{2(L+1)}\left [\frac{\sin(\frac{\pi}{2}(i-j))}{\sin(\frac{\pi}{2(L+1)}(i-j))}
  -\frac{\sin(\frac{\pi}{2}(i+j))}{\sin(\frac{\pi}{2(L+1)}(i+j))} \right]
\label{eq:initialC}
}
\par
It was shown recently that in the ground state there is a close connection between the 
entanglement entropies of the XY chain with $h=0$ and the TI chain \cite{IgloiJuhasz07}.
The consideration can be extended to the time-dependent case as sketched in Appendix A
and gives at criticality
\eq{
S_{XX}(L,t)= 2\; S_{TI}(L/2,t/2)
\label{eq:entropyXXTI}
}
One can see this relation clearly in numerical computations of $S$. For the spectra it means
that the same $\varepsilon_k$ occur in both systems, which one can also verify numerically.
In the critical case we will
therefore only present results for the XX model. These were obtained by determining $C_{ij}(t)$
in the subsystem by summing Bessel functions as in \cite{EislerPeschel07}.
For the non-critical case we will
consider the TI model. Here the procedure was to calculate the full correlation matrix 
${\bf{G}}(t)$ from ${\bf{G}}(0)$ for a sufficiently large total system using (\ref{eq:evolutionG}) 
and then to specialize to the subsystem. The size was chosen such that 
reflection phenomena in the larger subsystem could not yet enter the dynamics.

\section{Critical chains}

\subsection*{General behaviour}
\label{sec:crit_gen}

\begin{figure}[thb]
\center
\includegraphics[scale=.8]{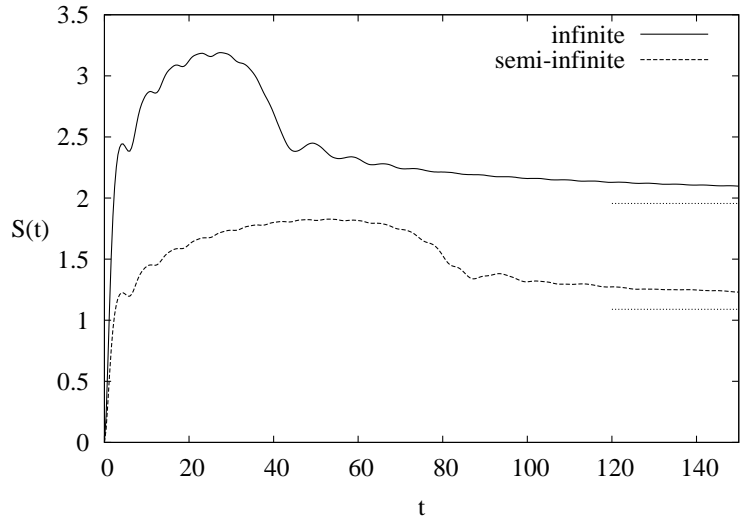}
\caption{Entanglement entropy for $L= 40$ and both geometries. The dotted lines
on the right are the corresponding equilibrium values.}
\label{fig:ent_dbl_semi}
\end{figure}

The general behaviour of the entanglement entropy is shown in Fig. 
\ref{fig:ent_dbl_semi} and can be described as follows. 
Starting from the value $S(0)=0$ (since the subsystems are
separated initially), it rises quickly to a broad plateau which extends to 
the times $t=2L$ in the semi-infinite and $t=L$ in the infinite
geometry, respectively. Afterwards it decreases slowly to its equilibrium value.
The characteristic times are those an excitation with the maximum
velocity $v_F=1$ needs in order to travel from the initial gap to the
end of the segment. In the semi-infinite case this involves a
reflection at the boundary, which gives $t=2L$.
\par
%
\begin{figure}[thb]
\center
\includegraphics[scale=.58]{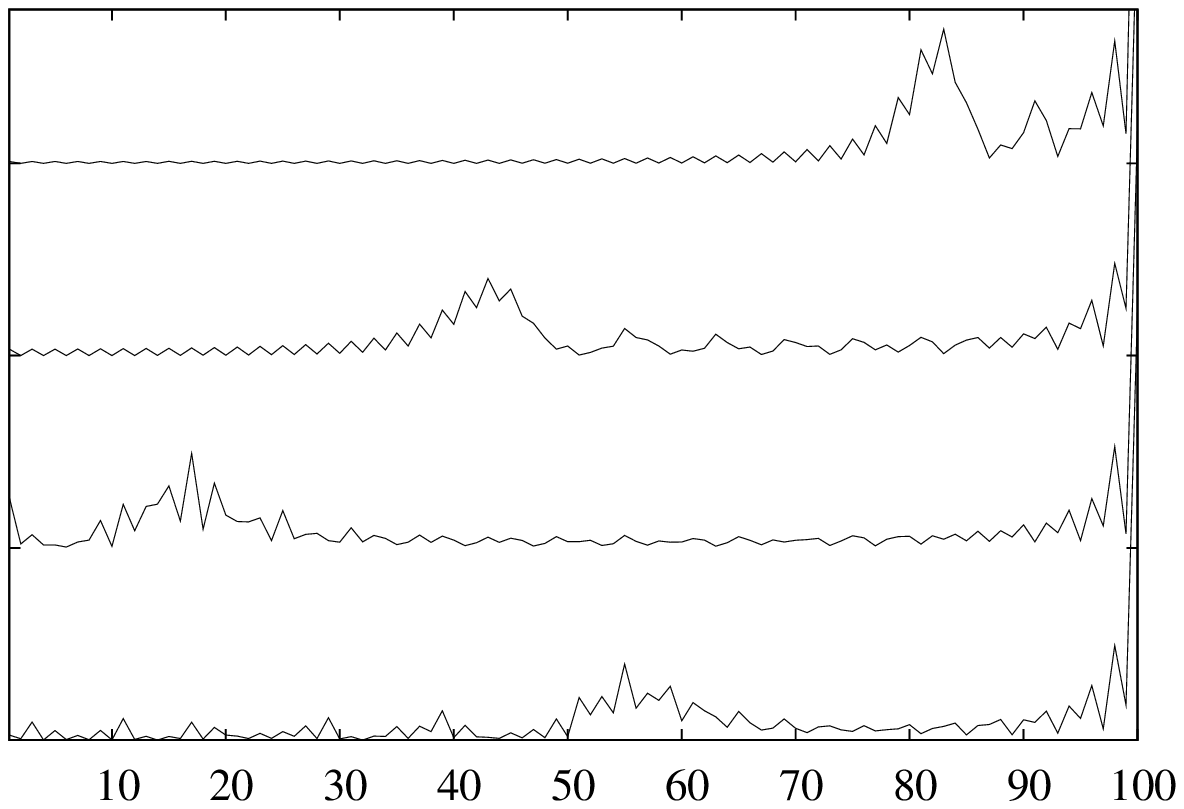}
\includegraphics[scale=.58]{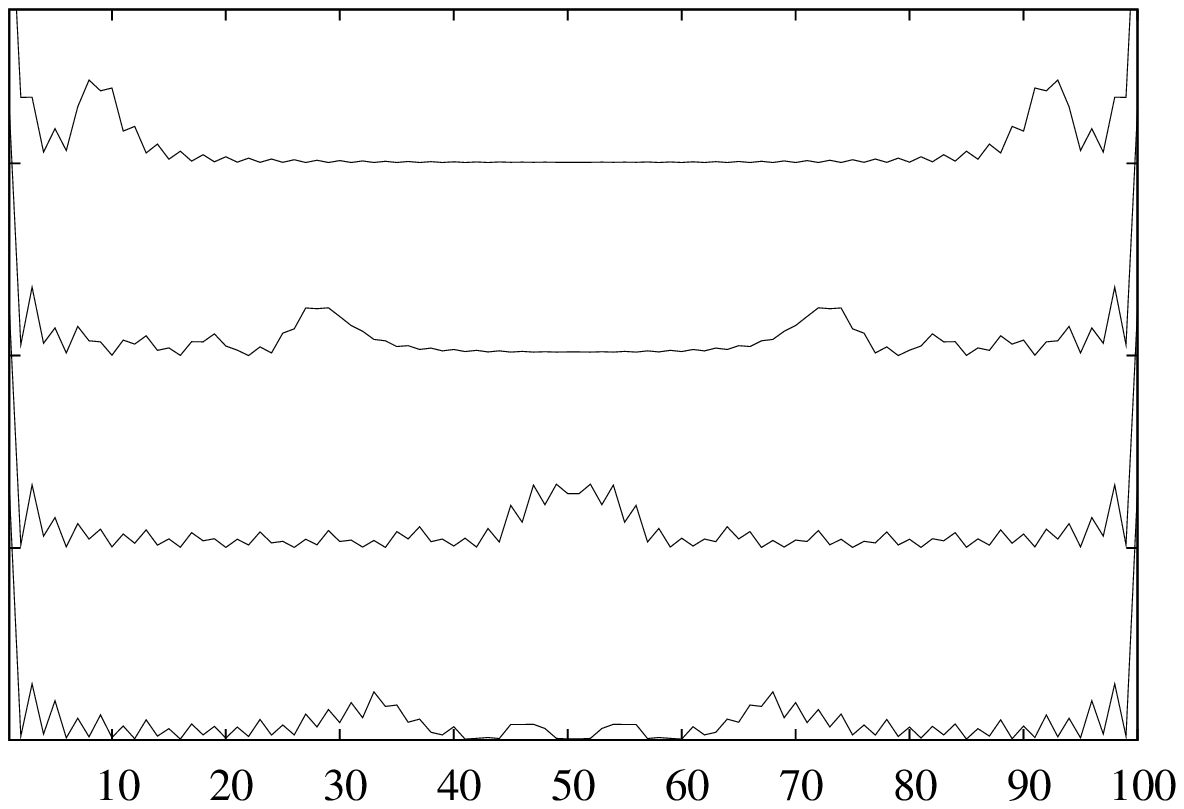}
\caption{Front propagation in the lowest single-particle eigenvector for $L=100$, with
time increasing from top to bottom.
Left: semi-infinite geometry, $t=20, 60, 120, 160$.
Right: infinite geometry, $t= 10, 30, 50, 70$.
Shown are the absolute squares of the amplitudes.}
\label{fig:fronts}
\end{figure}

%
Such a propagation phenomenon is clearly visible in the eigenvectors
of the single-particle states of $\rho$. Fig. \ref{fig:fronts} shows
the lowest one for four different times and both geometries.
In the semi-infinite case (left side) one sees a front propagating from
the initial defect towards the end where it is reflected. For the infinite 
geometry (right side) one has two fronts propagating inwards,
penetrating each other and continuing afterwards until they reach the
ends of the subsystem.
\par
%
\begin{figure}[thb]
\center
\includegraphics[scale=.75]{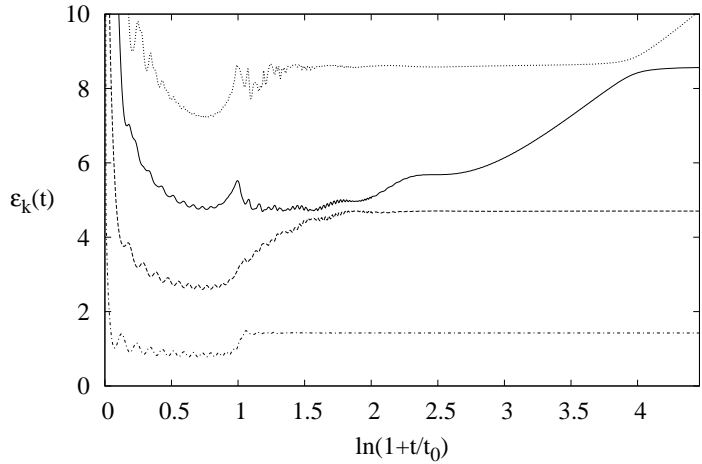}
\caption{Time evolution of the lowest four single-particle eigenvalues for
the semi-infinite geometry with $L = 40$. The parameter $t_0$ was chosen
such that $t=2L$ gives $1$ on the horizontal axis.}
\label{fig:eigvals}
\end{figure}

%
The evolution of the single-particle eigenvalues
is shown in Fig. \ref{fig:eigvals}. The horizontal scale has  been chosen
in order to make the features at both small and large times clearly visible.
The general picture is similar to the one that was 
found previously for one defect in an infinite system
\cite{EislerPeschel07}. In the beginning the $\varepsilon_k$ drop to some
transient values which produce the plateau in $S(t)$ and subsequently they
approach their equilibrium values. While this happens fast for the lowest 
$\varepsilon_k$, considerable deviations remain for long times among the
higher ones. These are the ``anomalous'' eigenvalues. The deviations, however,
are stronger than in \cite{EislerPeschel07} and there is a new feature in the 
form of certain intervals where the eigenvalue is roughly constant before
evolving again. We will come back to this phenomenon below.
%
%

\subsection*{Plateau region}
\label{sec:crit_smallt}

\begin{figure}[thb]
\center
\includegraphics[scale=.75]{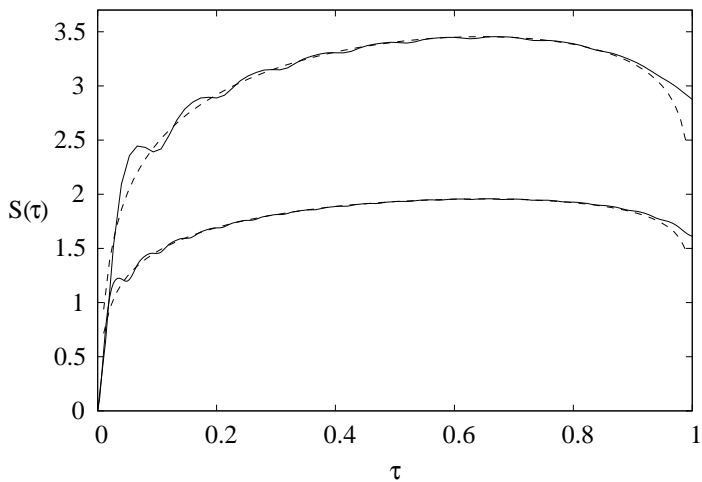}
\caption{Entropy plateau. Numerical results (solid) and
conformal predictions (dashed) for $L=60$. Upper curves: infinite case, $\tau =
t/L$. Lower curves: semi-infinite case, $\tau =t/2L$. }
\label{fig:cftfit}
\end{figure}

For the semi-infinite geometry, Calabrese and Cardy recently derived the 
following formula for $S(t)$ from conformal field theory  \cite{CC07}
\eq{S(t)=\frac c 6 \ln\left[ \frac{4L}{\pi}t
\sin \left(\frac{\pi t}{2L} \right) \right] + k'.
\label{entsemicft}}
Here $c$ is the central charge, equal to $1$ for the XX model, and
the cutoff parameter $\varepsilon$ appearing in the calculation
has been included in the constant $k'$. This formula is supposed to be valid
as long as the sine remains positive, i.e. for times up to $t = 2L$. 
The lower curves in Fig. \ref{fig:cftfit} show a comparison of numerical data 
with this prediction for a system wit $L=60$. There are deviations at the right 
and left end of the interval and the data shows additional oscillations, but 
overall the agreement is very good. Leaving the value of $c$ free, a fit between
$t = 10$ and $t = 110$ gives $c = 0.993$. Also the dependence on $L$ is described 
properly, i.e. curves for different $L$ collapse in reduced variables. Thus, in 
spite of an approximation involved deriving 
it \cite{CC07}, formula (\ref{entsemicft}) describes the lattice result very 
well.
\par
%
%
The infinite geometry was not treated in \cite{CC07}. However, as sketched in 
Appendix B, one can generalize the approach to this case by using an additional 
conformal mapping. The result after a rather long calculation is
\eq{S(t)=\frac c 3 \ln\left[ \frac{2L}{\pi}t
\sin \left(\frac{\pi t}{L} \right) \right] + k \,.
\label{entinfcft}}
Thus, compared to (\ref{entsemicft}), the prefactor has twice its value,
which is due to the two points of contact between subsystem and
surrounding, and $2L$ has been replaced by $L$. A comparison of this
formula to the numerical data is given by the upper curves in 
Fig. \ref{fig:cftfit}, again for $L=60$. The curves look very similar to 
the lower ones and the agreement is again very good. The same kind of fit
as before gives in this case $c = 1.007$
\par
The conformal field theory, which deals with particles of uniform
velocity $v=1$, does not give the oscillations seen in the numerics
and the relaxation at long times. Both are lattice effects connected
with the dispersion of the elementary excitations in the model.

\subsection*{Long times}
\label{sec:crit_larget}

\begin{figure}[thb]
\center
\includegraphics[scale=.55]{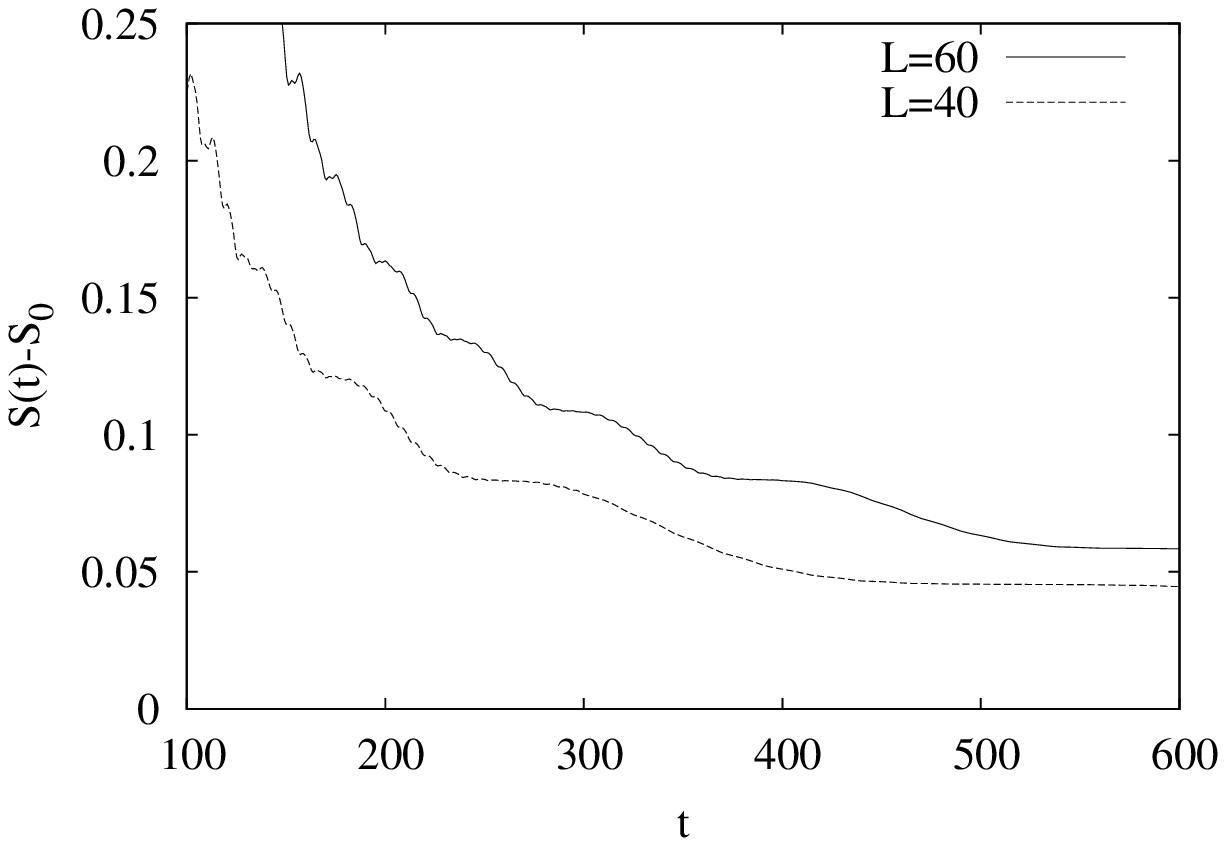}
\hskip0.5cm
\includegraphics[scale=.55]{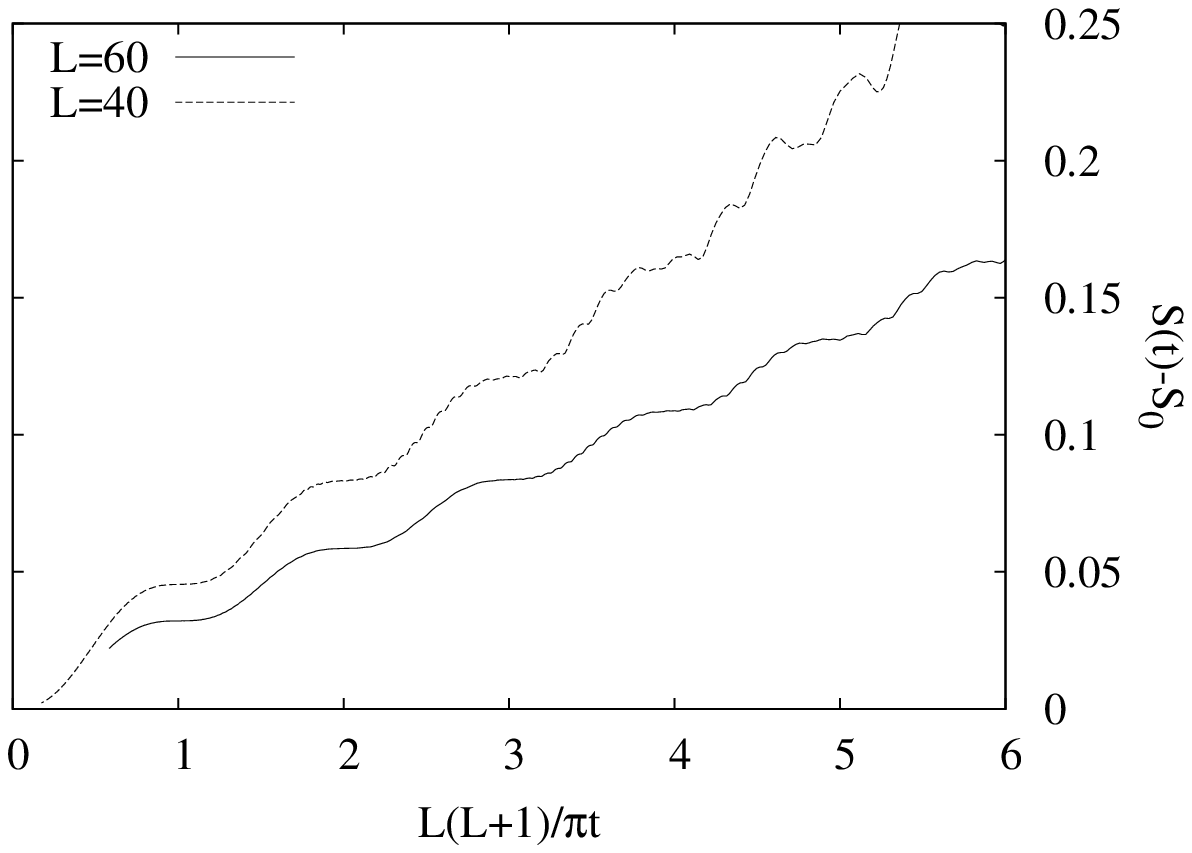}
\caption{Entanglement entropy at large times for the semi-infinite geometry. 
Left: $S$ vs. $t$. Right: $S$ vs. $1/t$. 
The equilibrium value $S_0$ is subtracted.}
\label{fig:entsteps}
\end{figure}

The behaviour at long times is more intriguing here than for a single
defect in an infinite chain. From Fig. \ref{fig:ent_dbl_semi} it might seem
that $S(t)$ simply decays towards its equilibrium value. Looking closer at
the semi-infinite case, however, one finds a clear step structure.
This is shown in Fig. \ref{fig:entsteps} for two values of $L$. These steps
become quite regular if one plots $S$ vs. $1/t$ as in 
the right part of the figure. Their height decreases with $L$ roughly according
to a power law $L^{-\alpha}$ where $\alpha \approx 0.90$. 
A comparison with the single-particle spectrum in Fig. \ref{fig:eigvals} shows
that the steps occur for those times where the anomalous eigenvalue
has the flat intervals mentioned previously. These are short for the earlier 
times and longer for the later times which gives the regular structure in $1/t$.
\par
%
\begin{figure}[thb]
\center
\includegraphics[scale=.55]{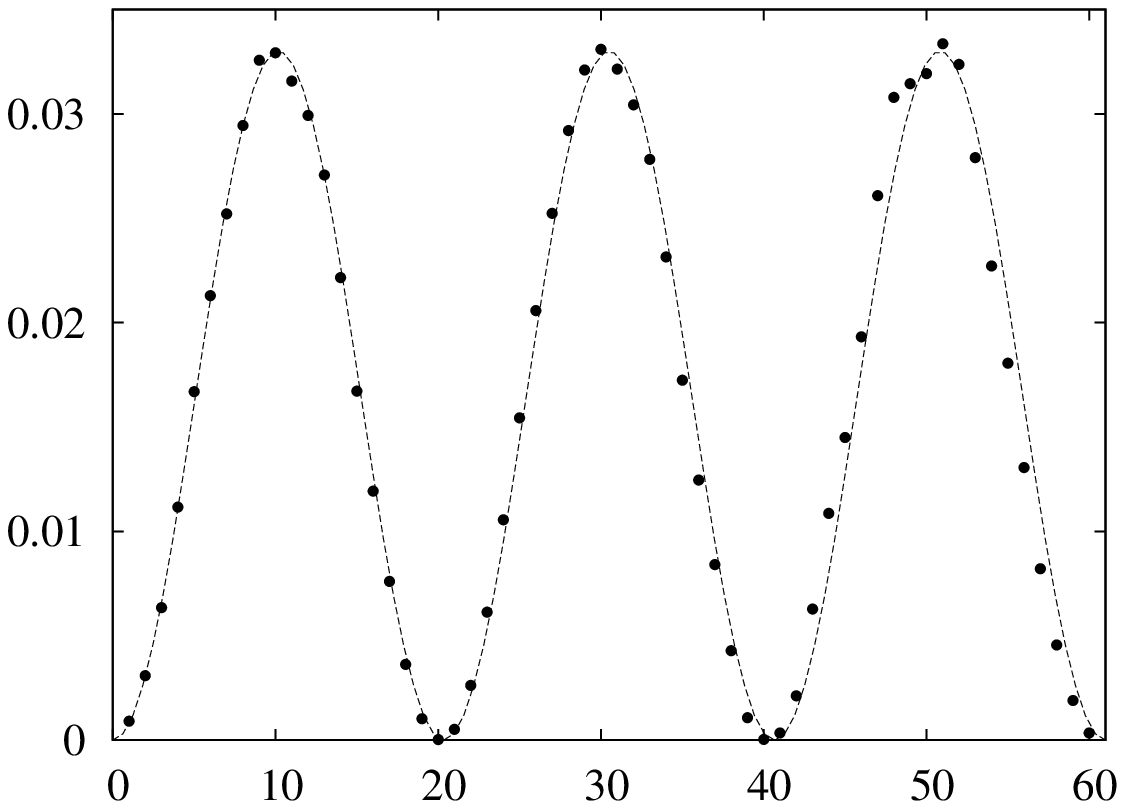}
\hskip0.5cm
\includegraphics[scale=.55]{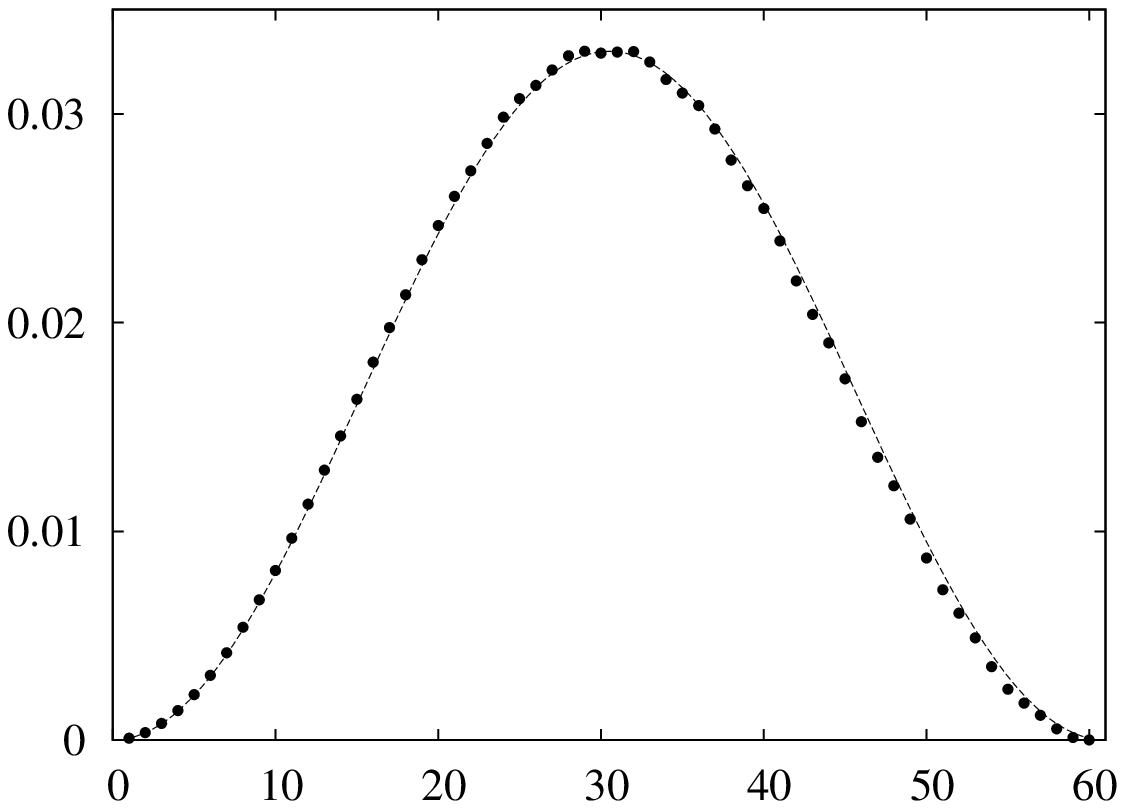}
\caption{Anomalous single-particle eigenvector for $L=60$. Left: $t = 390$. Right:
$t = 1160$. The dotted lines are fits with the function (\ref{evfit}).}
\label{fig:evecsin}
\end{figure}

%
If one examines the eigenvector of the anomalous single-particle state one finds
another surprise. For the flat intervals, it becomes remarkably simple. This is shown
in Fig. \ref{fig:evecsin} for $L = 60$ and two times related to the third-last and the 
last plateau, respectively. Plotted are the absolute squares of the amplitudes. 
%
These pictures suggest the functional form 
\eq{|\phi(j)|^2=A \sin^2(qj) \quad ; \quad q = \frac{\pi}{L+1}n,
\label{evfit}}
as for a standing wave in the isolated subsystem and the comparison shows that
this indeed fits the data very well. Moreover,
one can take these $q$-values and calculate the time an excitation with
$\omega_q=-\cos q$ created at the initial defect needs to travel
through the system. With $v_q=\sin q$ this is
\eq{T_n=\frac{L}{v_n} \approx \frac{L(L+1)}{\pi\;n}
\label{tnplateau}}
The times so calculated fit the plateau times in $\varepsilon_k(t)$
resp. $S(t)$ remarkably well. The expression also explains the scale used in the 
right part of Fig. \ref{fig:entsteps}. The plateaus then appear at integer values
of the variable.
\par
%
%
Taking all together, one can say that the discrete levels of the isolated subsystem
reappear in the time evolution. Since for long times the smallest $q$-values enter,
one is seeing here (in contrast to the fronts discussed earlier) the $\it{slowest}$
excitations in the subsystem. However, the selection of the $q$'s is not by interference
of a particle with itself, since (\ref{tnplateau}) gives only the time for travelling one
way. 
\par 
All considerations so far were for the semi-infinite case. In the infinite geometry, one
also finds such plateaus in the $\varepsilon_k(t)$ at times given by (\ref{tnplateau}) 
with $L$ replaced with $L/2$. The corresponding eigenvector also has the same features.
However, one does not observe plateaus in $S(t)$, because they are masked by the contribution
of another non-constant eigenvalue nearby.

\section{Non-critical chains}

\begin{figure}[thb]
\center
\includegraphics*[width=12cm]{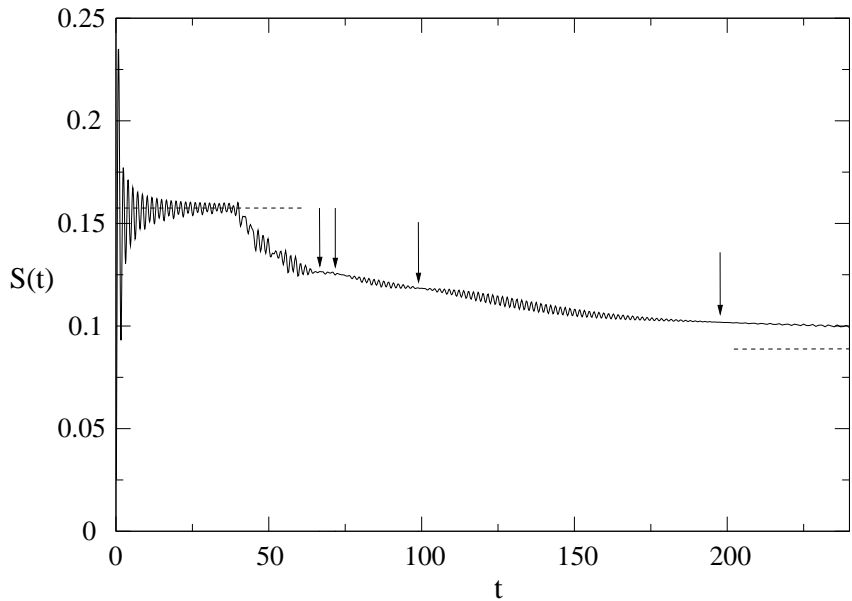}
      \caption{Entanglement entropy for the TI chain in the semi-infinite case
        with $h=2$ and $L=20$. The dashed horizontal lines indicate the plateau
        value and the equilibrium value, respectively. The arrows give the four
        longest times $T_n=L/v_n$ defined in the main text. 
 \label{fig:ent_noncrit_semi}  }
\end{figure}
\begin{figure}[thb]
\center
      \includegraphics*[width=8cm]{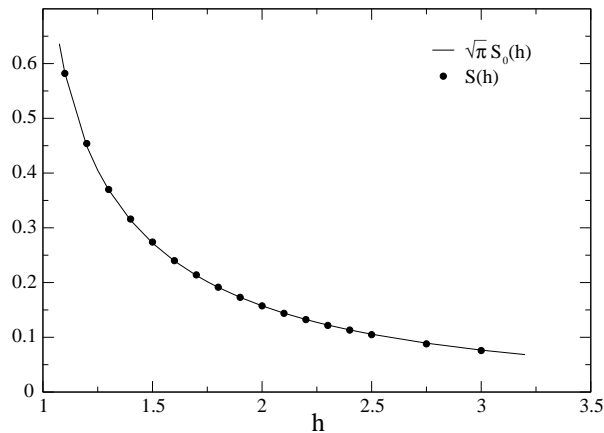}
      \caption{Plateau values of the entanglement entropy as a function of the field $h$.
        The line is the equilibrium value $S_0(h)$ multiplied by $\sqrt \pi$.
        \label{fig:plateau_noncrit}  }
\end{figure}

%
We now present results for the non-critical TI model in the disordered region $h>1$. The general
behaviour of $S(t)$ is shown in Fig. \ref {fig:ent_noncrit_semi} for the semi-infinite geometry
and $L=20$. This is a relatively small size but still much larger than the correlation length
$\xi=(h-1)^{-1}$ for the chosen value of $h$. As before, one sees an initial plateau which
extends up to $t=2L$. This is consistent with the velocity of the elementary excitations  
\eq{v_q=h\;\sin q /\omega_q
\label{velo_noncrit}}
which gives a maximum value $v=1$ for $q=\pi/2$ as in the critical case. The height of the plateau 
is independent of $L$ for $L \gg \xi$ and decreases monotonically as one moves away from the 
critical point.
At the shown value $h=2$ it is already rather small. Its detailed behaviour can be seen in
Fig. \ref{fig:plateau_noncrit}. One finds numerically that the plateau height is very accurately 
given by  $\sqrt{\pi}S_0$ where $S_0(h)$ is the equilibrium value of the entanglement entropy for 
large $L$ given analytically in \cite{Peschel04}. The relative deviation from this formula is 
less than $0.1 \%$.
\par
%
%

\begin{figure}[thb]
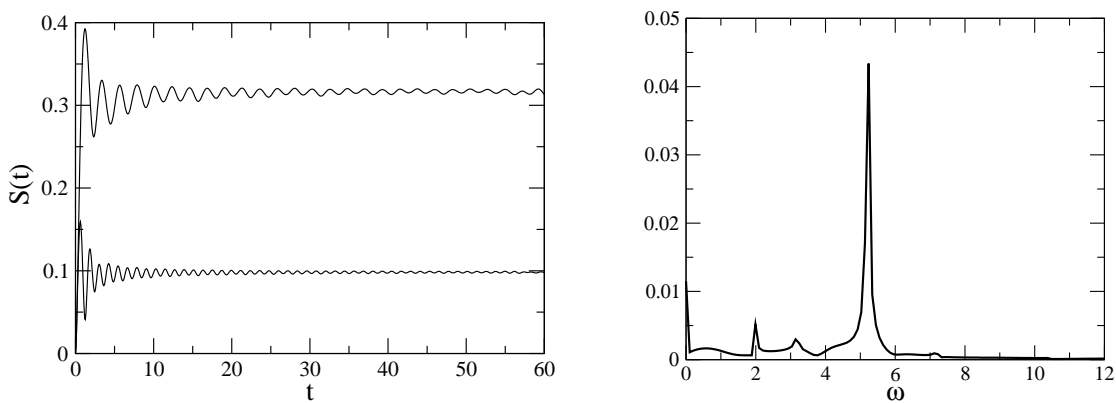

     \includegraphics*[width=7.3cm]{figs/early_times_L=30_h=1.4_h=2.eps}
\hskip1.0cm
     \includegraphics*[width=6.3cm]{figs/early_times_Fourier_L=30_h=2.6.eps}
      \caption{Left: Entanglement entropy in the plateau region for $L=30$,
        $h=1.4$ (top) and $h=2.6$ (bottom).
        Right: Power spectrum of $t[S(t)-\sqrt{\pi}S_0]$ for $h=2.6$.
        The main peak is located at $\omega=5.2$.
        \label{fig:oscill_noncrit}  }
\end{figure}

On top of the plateau one sees oscillations as in the critical case. Due to the small plateau
height, however, they are more prominent here. They reflect the lattice structure of the chain
and are shown in more detail on the left side of Fig. \ref{fig:oscill_noncrit}.
Their amplitude decays as $1/t$ while their period changes with the transverse field. 
From the Fourier spectrum on the right side of the figure one sees that they contain two main 
frequencies. These are given by $\omega_\pm = (h+1)\pm (h-1)$  
where $h+1$ and $h-1$ are respectively the upper and lower band edge of the dispersion curve 
(\ref{eq:dispersion}). They would appear in a stationary phase approximation for the integrals
in the rotation matrices ${\bf{R}}(t)$ for the time evolution. Of these, the larger frequency
$\omega_+ = 2h$ contributes with higher weight. Moreover, there is a peak in the spectrum
at zero frequency. Thus, to leading order $S(t)$ in the plateau region is described by
the expression
\eq{S(t)\simeq \sqrt{\pi}S_0(h)+\frac{1}{t}\left[A+B\cos(2ht) \right]
\label{eq:S_noncrit_ana}}
where $A$ and $B$ are constants depending a priori on $L$ and $h$.
%

%
For times longer than $2L$, Fig. \ref{fig:ent_noncrit_semi} shows a slow decay of the entanglement
entropy which is similar to the one in the critical case and the curve seems to approach the 
equilibrium value $S_0$. If one assumes that, the decay is approximately a power law $t^{\alpha}$
with an exponent near one. Due to the calculational procedure, the accessible times are smaller
here and limited to about 300. In this region, the spectrum of the single-particle eigenvalues
still deviates markedly from its asymptotic form. Only the lowest one has the value
one finds from the relation to corner transfer matrices \cite{PKL99,Peschel04}, namely
\eq{\varepsilon_1 = \pi I(k')/I(k) , \,\,\, k=1/h , \,\,\, k'=\sqrt{1-k^2}
\label{equ:eps_ctm}}
where $I(k)$ is the complete elliptic integral of the first kind, while the next two 
are irregular. Thus one cannot check the approach to equilibrium very well.
The curve for $S(t)$ shows again certain structures, but these are nodes between
oscillatory regions rather than steps. Nevertheless, as can be seen from Fig.
\ref{fig:ent_noncrit_semi}, they occur at the characteristic times $T_n=L/v_n$
calculated from the velocities (\ref{velo_noncrit}). The allowed values of $q$
in the non-critical case are not equidistant and lead to different velocities
near the upper and lower band edge.
Finally, one should note that the overall behaviour of $S(t)$ has a certain universal
character, since only one time scale, namely $L$, enters.
Thus one obtains an approximate collapse of the curves for different $L$ (but fixed $h$)
if one plots them as a function of $t/L$. This means that after times $t>2L$,
the decay is slower for larger subsystems. 
\section{Summary and discussion}

We have studied a special case of a local quench, namely the joining of initially separated
parts of a one-dimensional system, all prepared in their ground states. The quantity of interest
was the entanglement entropy between a finite piece and the remainder of the chain. We found
that $S$ behaves in a similar way as after the removal of a single defect from an infinite
system \cite{EislerPeschel07}. There is a transient regime where entanglement builds up 
rapidly and a plateau forms. Its length is determined by the velocity of the fastest particles.
In the critical case, its form is given by a universal logarithmic function and follows
from conformal field theory. For non-critical chains, 
it is rather flat and its height varies with the parameters of the system. At later times, there is
a slow decay towards the equilibrium entanglement in all cases. Thus one has a kind of overshooting
phenomenon.
\par
The critical case turned out to be particularly interesting for the semi-infinite geometry.
Then the final decay of $S$ proceeds in a step-like fashion with flat regions determined by
the travelling times of the slowest particles in the initially isolated subsystem. Thus, while
the gradual descent is generally a consequence of the dispersion in the velocities, it shows
here some memory of their initial discreteness. This is connected  with a special behaviour of
the anomalous single-particle eigenvalues of the reduced density matrix and their eigenvectors.
On that level, it seems rather general for this type of quench. However, to be observable in
the entanglement entropy, a proper eigenvalue structure is necessary. Although we can describe 
the phenomenon rather precisely, we have no complete picture of the mechanism. The same holds
for the simple formula giving the plateau height for the TI chain in terms of the equilibrium
entanglement. 
\par
The physical situation we have addressed, has been considered with a different focus in various
other papers. For example, in \cite{ARRS99} oppositely magnetized XX half-chains were brought into 
contact and the interest was in the magnetization profile. In a fermionic picture this corresponds
to charge transport via tunnel contacts as studied in \cite{Schoenhammer07}. 
Generalizations include the case of finite temperatures and a sandwich structure corresponding 
to our infinite geometry \cite{Ogata02,Asch06,PT05,PT07}. In \cite{Asch07} also the entanglement 
for large times was considered. However, for finite temperatures the entanglement entropy becomes 
extensive and does not measure the quantum properties we were studying here. Nevertheless, such an 
extension can put the zero-temperature features in a broader context.

\section*{Acknowledgement}
We would like to thank W. Aschbacher, P. Calabrese, J. Cardy, J. Eisert, K. Sch\"onhammer
and U. Schollw\"ock
for discussions.

\section*{Appendix A: Entanglement in XY and TI chains}

As pointed out recently \cite{IgloiJuhasz07}, there is a close connection between
the ground state
entanglement entropy of the zero field $XY$ chain and the TI chain. This was derived
by a detailed comparison at the level of the fermionic Hamiltonians and the correlation
matrices. Here we extend this to the non-equilibrium case using more general arguments.
\par
It has been known for a long time \cite{PeschelSchotte,Turban} that the zero field $XY$ 
Hamiltonian (\ref{eq:Hamiltonian}) can be split into a sum of two independent TI 
Hamiltonians by 
distinguishing even and odd lattice sites and using a dual transformation. If the couplings
$(1\pm\kappa)$ in (\ref{eq:Hamiltonian}) are denoted by $J_x$ and $J_y$ these are 
\eq{
H_{I,1}=-\frac{1}{2}\sum_{i=1}^{N/2-1} J_x \tau_i^{x,1}\tau_{i+1}^{x,1}-
\frac{1}{2}\sum_{i=1}^{N/2} J_y \tau_i^{z,1}
\label{eq:A1}
}
and
\eq{
H_{I,2}=-\frac{1}{2}\sum_{i=1}^{N/2-1} J_y \tau_i^{x,2}\tau_{i+1}^{x,2}
-\frac{1}{2}\sum_{i=1}^{N/2} J_x \tau_i^{z,2}
\label{eq:A2}
}
Here the $\tau$ operators are related to the $\sigma$'s via \cite{IgloiJuhasz07}
\begin{eqnarray}
\tau_i^{x,1}=\prod_{j=1}^{2i-1}\sigma_j^x\; ,
\qquad \tau_i^{x,2}=\prod_{j=1}^{2i-1}\sigma_j^y
\nonumber\\
\tau_i^{z,1}=\sigma_{2i-1}^y\sigma_{2i}^y\; ,
\qquad \tau_i^{z,2}=\sigma_{2i-1}^x\sigma_{2i}^x
\end{eqnarray}
One then has the decomposition
\eq{
H_{XY}=\frac{1}{2}\left[H_{I,1}+ H_{I,2}\right]
\label{eq:A3}
}
with $[H_{I,1},H_{I,2}]=0$.
If the $XY$ chain is cut at one point, the TI chains are also cut. Therefore, the initial
ground state of the $XY$ chain factorizes as
\eq{
\Phi_{XY}(0)=\Phi_{I,1}(0)\;\Phi_{I,2}(0)=
\Phi^1_{I,1}(0)\;\Phi^2_{I,1}(0)\;\Phi^1_{I,2}(0)\;\Phi^2_{I,2}(0) \; .
\label{eq:A4}
}
where the superscripts refer to the two subsystems. 
The evolution operator $U_{XY}(t)=e^{-itH_{XY}}$ factorizes as well, 
since $[H_{I,1},H_{I,2}]=0$
\eq{
U_{XY}(t)=U_{I,1}(t/2)\; U_{I,2}(t/2)
\label{eq:A5}
}
and the state at a later time is given by
\eq{
\Phi_{XY}(t)=\Phi_{I,1}(t/2)\; \Phi_{I,2}(t/2)\; .
\label{eq:A6}
}
Tracing over the unwanted degrees of freedom, one obtains the reduced density matrix as
\eq{
\rho_{XY}(t)=\rho_{I,1}(t/2)\; \rho_{I,2}(t/2)
\label{eq:A7}
}
and, restoring the length $L$, one has for the entanglement entropy
\eq{
S_{XY}(L,t)=S_{I,1}(L/2,t/2)+S_{I,2}(L/2,t/2)\; .
\label{eq:A8}
}
In the isotropic case, i.e. for the $XX$ chain, the two TI models are identical and critical
and (\ref{eq:A8}) gives (\ref{eq:entropyXXTI}) in Section 2.

\section*{Appendix B: Conformal mapping}

For the semi-infinite chain, the formula (\ref{entsemicft}) was derived in \cite{CC07}
within a path-integral approach for the quantum system. This leads to considering
a complex $(x,i\tau)$ plane, where $\tau$ is the imaginary time. In this plane,
one has a vertical line representing the boundary and a pair of half-infinite 
vertical lines representing the cut in the chain which exists up to $\pm i\varepsilon$. The 
entanglement entropy $S(t)$ is obtained from the expectation value of an operator
$\Phi_n(i\tau_0)$ in the gap between these half-lines after the analytic continuation 
$\tau_0 \rightarrow it$. To obtain this expectation value, one first simplifies the 
geometry by mapping the area to the right of the boundary line, where the path integral
is to be taken, onto the right half-plane. As explained in \cite{CC07}, this can
be done approximately by the combination of a Joukowsky transformation, a reflection
at the unit circle and a logarithmic map. The resulting transformation 
is analytically invertible which is needed for the application of the conformal formulae.
\par
%
\begin{figure}[thb]
\center
\includegraphics[scale=.6]{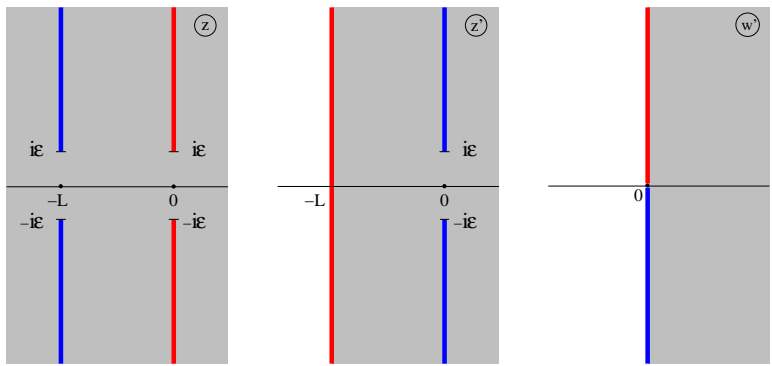}
\caption{Conformal mapping of the infinite geometry with two cuts (left) to the 
semi-infinite one with one cut (center) and onto the right half-plane (right). The
path integrals are taken in the dark areas. The colours show how the various
lines transform.}
\label{fig:confmap}
\end{figure}
%
The infinite chain with initially two cuts can be related to this problem in the 
following way. One has to consider in this case a complex plane with $\it{two}$ pairs of 
half-infinite vertical lines and the path integral is taken everywhere. The situation is
shown in Fig. \ref{fig:confmap} on the left. The transformation
\eq{
z' = \frac{1}{2}(\sqrt{z^2+\varepsilon^2} -z)-L)
\label{eq:B1}
}
which is a Joukowsky transformation as in \cite{CC07}, followed by a reflection, an
inversion and a translation by $L$ then maps this approximately onto the shaded region of the 
$z'$-plane shown in the center of Fig. \ref{fig:confmap}. This is exactly the geometry for the
single cut and one can use the previous procedure to map it to the right $w'$ half-plane.
The only additional feature is that the entanglement entropy follows here from a two-point 
function $\langle \Phi_n(i\tau_0)\Phi_n(i\tau_0-L) \rangle$. From its form in the $w'$-plane
\cite{CC07}
\eq{
\langle \Phi_n(w'_1)\Phi_n(w'_2)\rangle = c_n^2 \left[\frac{|w'_1+w'_2|^2}
{|w'_1-w'_2|^2 \;|2Re w'_1|\; |2Re w'_2|} \right]^{x_n}
\label{eq:B2}
}
where $x_n=c(n-1/n)/12$ one obtains it in the $z$-plane by the usual conformal transformation
formula.
\eq{
\langle \Phi_n(z_1)\Phi_n(z_2)\rangle = \left[ \; \left|\frac {dw'}{dz}\right|_1 \left| \frac 
{dw'}{dz} \right|_2 \;\right]^{x_n} \langle \Phi_n(w'_1)\Phi_n(w'_2)\rangle
\label{eq:B3}
}
After a lengthy calculation and analytical continuation, assuming $t,L \gg \varepsilon$, this 
leads to
\eq{
\langle \Phi_n \Phi_n \rangle =  c_n^2 \left[ \left(\frac{\pi}{2L}\right)^2
\left(\frac{\varepsilon}{t}\right)^2 \frac{1}{\sin^2(\pi t /L)} \right]^{x_n} 
\label{eq:B4}
}
Here the first two factors in the brackets come from the derivatives $dw'/dz$ while the 
sine arises from the denominator in (\ref{eq:B2}). The entanglement entropy is the negative
derivative with respect to $n$ at $n=1$ which gives (\ref{entinfcft}) in section 3. 

\section*{References}

%
%

\providecommand{\newblock}{}

\end{document}